\documentclass[usenatbib]{mn2e}
\usepackage{natbib}
\usepackage{hyperref}
\usepackage{psfig}
\usepackage{epsfig}
\usepackage{pdflscape}
\newif\ifAMStwofonts

\def\sqiglt{\hbox{\rlap{\lower.55ex \hbox {$\sim$}}\kern-.05em \raise.4ex \hbox{$<$}\,}}
\def\sqiggt{\hbox{\rlap{\lower.55ex \hbox {$\sim$}}\kern-.05em \raise.4ex \hbox{$>$}\,}}
\def\til{\ensuremath{\sim\,}}

\newcommand{\tim}[1]{\ensuremath{\times 10^{#1}}}
\def\deg{\ensuremath{^{\circ}}}

\def\cms{\ensuremath{$cm$^{-2}}}

\def\swift{\emph{Swift}}

\def\t0{\ensuremath{T_{0}}}

\def\arcmin{\ensuremath{^\prime}}
\def\nu{\ensuremath{\nu}}

\def\toa{\ensuremath{t_{\rm oa}}}

\title[Swift follow-up of ALIGO/AVirgo triggers]{Optimisation of the \swift\ X-ray follow-up of
Advanced LIGO and Virgo gravitational wave triggers in 2015--16}

\author[Evans et al.]{P.A. Evans$^1$\thanks{pae9@leicester.ac.uk}, J.P. Osborne$^1$, J.A. Kennea$^2$,
S. Campana$^3$ P.T. O'Brien$^1$,
\and N.R. Tanvir $^1$, J.L. Racusin$^4$, D.N. Burrows$^2$, S.B. Cenko$^{4,5}$, N. Gehrels$^4$
\\
$^1$Department of Physics and Astronomy, University of Leicester, Leicester, LE1 7RH, UK
\\
$^2$Department of Astronomy and Astrophysics, Pennsylvania State University,
525 Davey Lab, University Park, PA 16802, USA
\\
$^3$INAF -- Osservatorio Astronomico di Brera, via E. Bianchi 46, I-23807, Merate, Italy
\\
$^4$Astrophysics Science Division, NASA Goddard Space Flight Center, Mail Code 661, Greenbelt, MD 20771, USA
\\
$^5$Joint Space-Science Institude, University of Maryland, College Park, MD 20742, USA
}

\date{Accepted -- Received --}

\pagerange{\pageref{firstpage}--\pageref{lastpage}}
\pubyear{2015}

\begin{document}

\maketitle

\label{firstpage}

\begin{abstract} 
One of the most exciting near-term prospects in physics is the potential 
discovery of gravitational waves by the advanced LIGO and Virgo detectors. To maximise
both the confidence of the detection and the science return, it is essential
to identify an electromagnetic counterpart. This is not trivial, as the events are
expected to be poorly localised, particularly in the near-term, with error regions
covering hundreds or even thousands of square degrees. In this paper we discuss the prospects for finding an X-ray counterpart
to a gravitational wave trigger with the \swift\ X-ray Telescope, using the assumption
that the trigger is caused by a binary neutron star merger which also produces a short gamma-ray burst.
We show that it is beneficial to target galaxies within the GW error region, highlighting
the need for substantially complete galaxy catalogues out to distances of 300 Mpc. 
We also show that nearby, on-axis short GRBs are either extremely rare, or
are systematically less luminous than those detected to date. We consider the prospects
for detecting afterglow emission from an an off-axis GRB which triggered the GW facilities, finding that
the detectability, and the best time to look, are strongly dependent on the characteristics
of the burst such as circumburst density and our viewing angle.
\end{abstract}


\section{Introduction}
\label{sec:intro}

In the last quarter of 2015 the Advanced LIGO observatory (ALIGO; 
\citealt{Harry10}) is expected to begin its first science run, spending 
three months searching for gravitational waves (GW). In 2016 it will
be joined by Advanced Virgo (AVirgo; \citealt{Acernese15}), and perform a six-month long run
\citep{Aasi13}.
Electromagnetic (EM) follow-up of GW triggers is essential to maximise the 
scientific information gleaned from any such detection, as well as 
providing a useful confirmation of the GW signal. 
To this end, much effort is currently being expended both in predicting what EM signal 
will accompany a GW event (e.g.\ \citealt{Metzger12,Andersson13}), and in planning 
observational follow-up strategies (e.g.\ \citealt{Nissanke13}). Of course, this
depends on what gives rise to the GW signal in the ALIGO/Virgo bandpass. While various phenomena (e.g.\ core-collapse
supernovae, magnetar flares) are expected to give rise to GW signals, the most promising
phenomenon is believed to be the merger of two stellar mass compact objects such
as neutron stars (NS) or black holes. The final stages of such a merger are expected
to emit a GW `chirp' that could be detected by ALIGO/AVirgo, if the event is
near enough to Earth. The predicted rates of such mergers that will be detectable
with ALIGO/AVirgo is highly uncertain, with estimates (e.g.\ \citealt{Abadie10,Coward12}) covering up to 4 orders of magnitude!
Electromagnetically, such a merger is accompanied by
a range of emission (see \citealt{Metzger12}) including that of a
short gamma-ray burst (sGRB). These are short ($\sqiglt2$ s) but 
extremely intense ($L\til10^{51}$\ erg s$^{-1}$) flashes of high energy 
radiation, followed by a panchromatic `afterglow' from the shocked
circumstellar material; see \cite{Berger14} for a review. The marriage of 
EM and GW signals from an sGRB is extremely valuable as the mass, distance and binary
inclination information potentially available from the GW signal complements
the luminosity, redshift and duration measurements available from EM observations to constrain
accurately the energetics and hence physics of the merger process. These measurements
also have the potential to constrain cosmology, through combining GW distance information
with spectroscopic redshifts \cite[e.g.][]{Schutz86,Nissanke10}.
Additionally, a precise EM location should permit
the identification of the host galaxy and source environment.

In this paper we consider how best to respond to GW triggers with the 
\swift\ satellite \citep{GehrelsSwift}. We are only considering the case 
that that the GW event is accompanied by an sGRB, and we assume that 
the sGRB arose from an NS-NS merger, i.e.\ we do not consider
NS-BH mergers. This is primarily because the input to our work
are the GW simulations of \cite{Singer14}, who simulated the GW signal
from NS-NS systems. The main consequence of this is that we may be underestimating
the typical distance of the GW triggers, since NS-BH binaries can be 
detected out to greater distance than the NS-NS binaries \citep[e.g.][]{Aasi13}.
\swift\ was designed 
to study GRBs and, at the time of writing, the on-board Burst Alert 
Telescope (BAT; \citealt{BarthelmyBAT}) has detected nearly 1,000, 
including \til85 sGRBs\footnote{Defined here na\"{\i}vely as simply 
those with $T_{90}\le2$ s.}, and localised them to within 3\arcmin\ at 
trigger time. Of those bursts which were promptly observed by the narrow 
field, 0.3--10 keV X-ray telescope (XRT; \citealt{BurrowsXRT}) almost 
all (94\%\ of all GRBs, 80\%\ of short GRBs) have been detected and 
around 40\%\ were detected by the UV/optical telescope (UVOT; 
\citealt{RomingUVOT}). \swift-BAT has a peak sensitivity at a 
significantly lower energy than \emph{CGRO}-BATSE and \emph{Fermi\/}-GBM 
and -LAT, which disfavours the detection of sGRBs (the fraction of 
\swift-triggered GRBs that are short is much lower than for these other 
missions). However, very nearby events, such as expected
as GW detections, are likely to be bright in gamma-rays
and largely unaffected by this bias. A more significant
issue is the BAT's constrained field of view (2~sr, 
i.e. \til$\frac{1}{6}$ of the sky) which means that it is rather
unlikely that an sGRB which triggers ALIGO/AVirgo will also trigger the BAT
(this is without considering the question of whether typical mergers are
likely to produce bright sGRBs, which we discuss in Section~\ref{sec:howbright}).
Therefore, in order to localise the sGRB with \swift\ we will need to rely on 
follow-up with the narrow-field instruments. In this work we focus on 
the XRT, since it has detected almost every GRB it has looked for; and, 
as we will demonstrate in Section~\ref{sec:falsePos}, the rate 
of`contaminating' X-ray transients unrelated to the GW trigger is much 
lower than at optical wavelengths.

While the BAT GRB localisation fits easily inside the XRT field of view (radius 
12.3\arcmin); for GW events the error 
region will likely cover many hundreds of square degrees (e.g. 
\citealt{Nissanke13,Singer14,Berry15}). \swift\ not only has the
ability to slew rapidly, but it can also autonomously perform multiple
overlapping `tiled' observations \citep{Evans15}, allowing it to cover
large error regions, although this has not yet been attempted for
regions as large as expected from ALIGO/AVirgo. \swift\ has previously
observed the fields of two GW triggers \citep{Evans12}, but these observations
were extremely limited, covering less than one square degree between the two triggers.
\cite{Kanner12} explored ways of observing the large GW error regions with XRT, and suggested that, initially,
XRT should observe a subset of nearby galaxies in the GW error region. When the 
GW horizon distance increases, they suggest observing 
a series of tiled pointings, of 100 s each. Despite this work, it 
is not obvious what the \emph{optimal} means of observing the ALIGO/Virgo error region with the
\swift-XRT is, both in terms of where to look, and how long to look for.
In this paper we use end-to-end simulations to explore the optimal approach 
to following up GW triggers with \swift-XRT, under the assumption that the 
underlying event is a NS-NS merger accompanied by an sGRB.

\section{How to identify the X-ray afterglow}
\label{sec:challenge}

As noted by authors such as \cite{Metzger12}, when responding to
triggers such as those from ALIGO/AVirgo, we do not simply need to detect the EM
counterpart, but also to identify it as such, from among the
unrelated sources detected. \cite{Evans15} proposed 
two ways  of distinguishing an X-ray GRB afterglow from
unrelated, uncatalogued sources using \swift: brightness or fading behaviour. If
a source is bright enough that it should have been catalogued, but
is not, then it is by definition some form of transient phenomenon, and
could be the afterglow. Alternatively, GRB afterglows are known to fade
on short timescales, therefore an uncatalogued and fading object is also
a good candidate afterglow, even if it is not above the catalogue limit.
However, GRBs are not the only transient phenomena in the X-ray sky, and when 
observing the large GW error regions we need to consider the possibility of 
coincidentally detecting a transient, unrelated to the GW signal, which we now do.

\subsection{False alarm rates}
\label{sec:falsePos}

To investigate the rate of bright X-ray transients, we selected
all GRB datasets (apart from stacked images) from the 1SXPS catalogue
\citep{Evans14}, which gives position and flux (among other things) for all point
sources observed by XRT up to 2012 October. We limited ourselves to the GRB fields as, in these
cases, which X-ray source was the target of the observation is known, so cannot be mistaken for
a serendipitous or unrelated object. Within each of these fields, we searched for
X-ray sources which were not the GRB, were not catalogued X-ray
emitters\footnote{Unless the catalogue entry was from one of the earlier XRT catalogues,
i.e.\ detected in the same dataset as in the 1SXPS catalogue.}, had a detection
flag of \emph{Good, Reasonable} or \emph{Poor}, and were brighter than a
3-$\sigma$ upper limit from the \emph{Rosat\/} All Sky Survey (RASS; 
\citealt{Voges99}), calculated at the location of the
source (see Section~\ref{sec:simulate} for details of the RASS upper
limit determination). That is, we selected all objects which would trigger
our `bright source' criterion but lie coincidentally in the XRT field of view. We found 32 such objects, in a total of
52.6 Ms of observing time, covering \til95 square degrees on the sky. 

To infer the rate of these
transients requires knowledge of their duration. If they are all short-lived (compared
to the typical duration of an XRT exposure, \til 500 s) then the rate at which X-ray transients `turn-on' is
1 per 1.64 Ms per XRT field of view (0.12 square degrees). 
The observing strategies explored later in this paper cover XRT exposure times of 0.006 --1 Ms 
per GW trigger; with those defined as `best' (see Section~\ref{sec:res2015}--\ref{sec:2016})
requiring 0.03--0.18 Ms per trigger. So, for each GW trigger we follow with XRT, this predicts a \til2--10\%\ chance of detecting a bright
X-ray transient unrelated to the GW trigger. This is much lower than the
expected rates of optical transients, which could be as high as hundreds of
events per GW trigger (e.g.\ \citealt{Nissanke13}, although see, e.g., \citealt{Cowperthwaite15} for
ways to mitigate this). It is also an upper limit, since the
1SXPS observations from which it was derived are of sufficient exposure to detect
all sources above the RASS limit, whereas, for most strategies explored in Section~\ref{sec:simulate}
the exposures are much shorter than this, so most of
the 32 bright sources in 1SXPS would not be detected.

On the other hand, if the transients last long enough to all be `active' at
the same time (\til5--10 years), then we have determined that transients exist
on the sky with a density of 0.3 per square degree. The strategies we will explore in this work
cover 6--216 square degrees, thus we would predict \til2--65 transients unrelated to the 
GW event to be detected by our observations. The reality is likely somewhere between these two values.
The \swift\ data do not allow us to place strong constraints on the duration of the 32 transient events
as, in all but 9 cases, the transient was either already above the RASS limit when XRT first observed it,
or still above the limit when XRT stopped observing it. For the 9 cases where both
the start and end of the transient phase were seen by XRT, the median duration is \til2\tim{5} s.
This is short compared to the time interval over which the 32 transients were detected (250 Ms),
so we suggest that the false positive rate is likely to be closer to the first value (one per XRT field of view
per 1.64 Ms) than the second one (0.3 per square degree).

\cite{Kanner13} have also investigated the expected rate of X-ray transients, based on
\emph{XMM-Newton} Slew Survey data. It is difficult
to directly compare their results with ours, since they used different selection criteria
(using only the 0.2--2 keV band, selecting sources with a flux 10$\times$\ the RASS 2-$\sigma$ upper limit,
and only those associated with known galaxies) which will detect fewer transients than our approach.
They also only gave the spatial density of transients, with no temporal information. However,
they predict that the probability of XRT detecting a transient unrelated to a GW trigger in observations covering
35 deg$^2$ and of 100 s each, is \sqiglt1\%. If we assume one transient per XRT field of view, per 1.64 Ms, we
predict 1.8\%; these numbers are reasonably compatible, given the above differences in measurements, and supports our
belief in the lower of the two false positive rates we have deduced.

To determine the rate at which fading X-ray sources unrelated to the GW event will be detected is more
complicated, since it depends heavily on the observing strategy
employed. However, we can gain an estimate of the rate from
\cite{Evans15}. In response to 20 neutrino triggers from IceCube, they
detected a total of 72 sources in their initial XRT observations, of
which only one initially appeared to be fading (further observations showed that is
was not related to the neutrino trigger). Therefore we can place a crude
estimate that \til1/72=1.4\% of objects detected by XRT but not related to the GW trigger
will show fading behaviour. 

These considerations demonstrate that, if a candidate counterpart to the GW event is detected in
X-rays, the probability of it being unrelated to the GW trigger is relatively small.
Additionally, since \swift\ is in space with a 96-min orbit, and only a 46\deg-radius
exclusion zone around the Sun and 22\deg\ zone around the Moon, it can observe a much larger fraction of
the sky (and hence GW error region) than a ground-based facility. This shows that 
\swift-XRT is a potentially valuable resource in searching for the EM counterpart to a GW trigger.

\subsection{Feasibility and strategy}

The size of the GW error regions (hundreds of square degrees) makes observing them
a challenge for the narrow-field XRT on \swift; therefore it is necessary to observe a large
number of `fields': that is, distinct (but possibly overlapping)
locations on the sky. To determine the optimal strategy for observing
these fields there are two chief questions which need investigating. The
first is whether to select XRT fields to cover the raw 
error region produced by ALIGO, or to target specific locations within 
it. With its modest field of view, XRT cannot in most cases, expect to 
cover the entire GW error region within a reasonable time after the 
trigger. The simplest approach is, if we can observe $N$ XRT fields, to 
select the $N$ fields which enclose the most probability (this is the 
concept of `searched area' in \citealt{Singer14}). However, sGRBs are 
expected to exist in or near to galaxies \citep[e.g.][]{Fong10,Tunnicliffe14}, therefore it may actually be 
better to first convolve the ALIGO probability map with an appropriate 
galaxy catalogue, and observe the most probable $N$ fields after this 
convolution: this is the approach which was followed in earlier \swift\ 
follow-up of LIGO-Virgo triggers \citep{Evans12}, and was advocated by \cite{Kanner12} 
and \cite{Nuttal10} for nearby GW events. This approach has the advantage 
of dramatically reducing the area that has to be covered, and therefore 
reducing the average time from the GRB to when XRT observes it. On the 
other hand, galaxy catalogues are not complete, and therefore some 
fraction of the time, this approach will mean discarding the XRT field 
which actually contains the GRB.

The second question is for how long each field is observed. Longer
observations\footnote{Here, and throughout, by `observation' we mean a
single exposure of a single field.} allow fainter sources to be
detected, however they also increase the time it takes to reach the
field which contains the GRB. Since GRB X-ray afterglows fade, this increased delay between the
GRB and the observation may make the GRB undetectable. Also, the longer
the exposure, the more unrelated sources we expect to detect,
making identifying the GRB a greater challenge (if we have to
re-observe the sources to look for fading, then the required observing
time scales with the number of unrelated uncatalogued sources).

\subsubsection{Galaxy Catalogues}
\label{sec:galcat}

To investigate the first question -- whether or not it is better to preferentially observe
fields containing known galaxies -- we require a galaxy catalogue to use in our
simulations. In this work we used the Gravitational Wave Galaxy Catalogue (GWGC; 
\citealt{White11}): a compilation of various pre-existing catalogues.
This gives the position, distance, size and
orientation of galaxies out to 100 Mpc. The ALIGO probability maps
(Section~\ref{sec:simulate}) are in {\sc healpix} format
\citep{Gorski05}, therefore we created a {\sc healpix}-format mask,
based on the GWGC. Since NS are often given a `kick' at the moment of
their birth, and indeed sGRBs are often found significantly offset
from their host \citep[e.g.][]{Fong10,Church11}, we added a 100 kpc halo to each
galaxy further than 5 Mpc (for galaxies closer than this, the angular
projection of a 100 kpc halo covers an unreasonably large fraction of
the sky, but the fraction of binary NS mergers which occur within this distance is negligible);
all pixels in the {\sc healpix} mask that lay within a galaxy
or its halo were set to 1, the remaining pixels were set to 0.

The GWGC catalogue is not complete, however an estimate of its
completeness (based on $B$-band luminosity) is given in fig.~5 of
\cite{White11}. Based on this figure we have tabulated the approximate
completeness of the catalogue as a function of distance in
Table~\ref{tab:gwgccomplete}. We interpolated within this table to
determine the completeness of the GWGC at various distances throughout
this paper. This completeness estimate is based on $B$ magnitude, which
is likely to be a reasonable indicator of sGRB rate since, as noted by
\cite{Fong13}, sGRB ``demographics are
consistent with a short GRB rate driven by both stellar mass and star formation.''
Therefore, the published GWGC completeness
will give a useful indication of the value or otherwise of using galaxy catalogues
in the follow-up of GW triggers.

A drawback to the GWGC is that it only extends to 100 Mpc, whereas in the 2016 science
run, when both ALIGO and AVirgo are active, we expect to detect GW events from beyond this horizon
\citep{Singer14}. While alternative catalogues exist which do extend beyond 100 Mpc, e.g.\ the 2MASS Photometric Redshift Catalog
(2MPZ; \citealt{Bilicki14}) -- which  contains photometric redshifts for galaxies detected
in the 2MASS infrared survey, with a median of $z=0.08$ (\til 300 Mpc) -- the completeness of
these has not been quantified, and the redshifts are photometric, so we have not used these in our simulations.

\begin{table}
\begin{center}
\caption{Completeness of the GWGC, based on fig.~5 of \protect\cite{White11}.}
\label{tab:gwgccomplete}
\begin{tabular}{cc}
\hline
Distance & Completeness \\
(Mpc) & (\%) \\
\hline
$\leq$40 & 100 \\
50 & 70 \\
60 & 65 \\
70 & 65 \\
80 & 60 \\
90 & 58 \\
100 & 55 \\
$>$100 & 0$^a$ \\
\hline
\end{tabular}
\end{center}
\medskip
$^a$ The GWGC only includes galaxies within 100 Mpc, hence the sudden cut-off.
\end{table}

\section{Simulation procedure}
\label{sec:simulate}

In order to simulate the \swift-XRT follow-up of sGRBs which trigger ALIGO/AVirgo, and explore the impact
of different observing strategies, we used the GW simulations of \cite{Singer14}. These authors simulated the GW emission
from a large number of NS-NS mergers distributed homogeneously in space,
produced the ALIGO/AVirgo signals which would be detected from these and applied the 
data analysis routines to those signals. This was done for the expected ALIGO configuration in 2015, and 
for the ALIGO-AVirgo configuration (and duty cycles) expected in 2016.
For the mergers which were detected by this process, the probability map and the input binary system
parameters, are available online\footnote{http://www.ligo.org/scientists/first2years/}.
We are only interested in the mergers which are accompanied by sGRBs, for which the 
jet of outflowing material is directed towards the Earth. 
The opening angle of sGRB jets is not well known \citep[e.g.][]{Margutti12,Fong14}, so in order to allow a good number of 
simulations, we took a value at the wide end of those predicted, and selected
from the \cite{Singer14} simulations all 
those with a binary inclination\footnote{While binary inclination angle is 
normally given in the range 0\deg--90\deg, the GW simulations
differentiate between clockwise and anticlockwise rotations, allowing
$i$ to be in the range 0--180; therefore we formally selected objects
with $i\leq30\deg$ and $i\geq150\deg$.} $i\leq30\deg$, resulting in \til200 simulated objects 
each year (corresponding to \til40\%\ of the mergers `detected' in the \citealt{Singer14} simulations).
\cite{Singer14} provided two probability maps -- all-sky maps where each pixel's value
is the probability that the GW event occurred in that pixel -- one map was derived rapidly using
an algorithm called {\sc bayestar}, the second created using a full
Markov Chain Monte Carlo (MCMC) analysis, which takes much longer to run
(typically \til109 hr according
to \citealt{Berry15}). We used the {\sc bayestar} maps, since these are available
for all of the triggers on the LIGO website (the MCMC map is not),
and because, due to their rapid creation and their similarity to the full MCMC results (for 2015, at least)
they are likely to be the maps that will be used for real XRT follow-up of a GW trigger.

As noted above, these simulations were distributed homogeneously in
space, whereas NS-NS mergers are expected to occur in or near to
galaxies, so we needed to modify the spatial distribution
of the simulated GW events to reflect this expectation. This can be
done relatively simply, because the shape of the GW error region is determined not by the celestial
location of the merger, but by the geocentric coordinates. This means that
we can shift the simulated mergers and probability map in RA without invalidating them,
provided we also also adjust the trigger time such that the GRB is at the same 
geocentric position as in the original simulations.

Therefore, for each GW simulation we determined the completeness of the GWGC 
at the distance of the GRB (Table~\ref{tab:gwgccomplete}).
For objects less than 100 Mpc away
(i.e. the distance range covered by the GWGC), a random number was
generated to determine whether this object should lie in a GWGC
galaxy\footnote{i.e.\ the random number, $R$, had a value in the range
$0\le R<1$. If the GWGC was, e.g. 70\%\ complete at the merger 
distance, then the merger was considered to be in a GWGC galaxy if 
$R\le0.7$.}. If it should, a galaxy was selected at random\footnote{A better approach
would be to weight the galaxies by luminosity, however since we cannot move the
\protect\cite{Singer14} simulations in RA this is not possible.}
from those 
within the GWGC which had a distance consistent with that of the 
simulated merger, and a declination within 3$\deg$ of the original LIGO 
simulation\footnote{It is not practical to require the shift to be purely in RA;
however a 3\deg\ declination shift is so small that the GW probability
maps are not invalidated by this shift.}. A position was randomly selected within this galaxy, with 
uniform probability through the galaxy, and then a kick of random 
magnitude (up to 100 kpc) and direction was applied to it. The
GW probability map was then rotated on the celestial sphere such that the GRB lay
at this position. For mergers where the random number determined that
it was not inside a GWGC galaxy, or mergers at a distance greater than
100 Mpc (i.e. not covered by the catalogue) no shift was applied. 

For each merger, regardless of whether we repositioned it, the day of 
the trigger was selected at random.  The time of day at which the merger 
would lie at the geocentric location used for the GW simulations on 
this day was determined: this was set to be the GRB trigger time. For 
the 2015 simulations, the times were all in a 90 day window starting on 
2015 September 1 (i.e.\ approximately the time expected to be covered by 
the science run); for 2016 we used a 180 day window starting in mid 
June. The only impact this has on the simulations is to set the positions of the Sun and Moon,
hence the regions of sky which \swift\ cannot observe.

As well as the GRB location, distance and error region just described,
the following two parameters were also needed for each simulation:

{\bf Time delay, $\tau_d$} -- the time interval in hours between the GW
trigger, and the commencement of \swift\ observations. Since the initial
GW probability map will be produced within minutes of the trigger
\citep{Singer14}\footnote{There may be an initial process of human
screening before the map is distributed, but this is still being debated,
so we have ignored it.}, we drew $\tau_d$ at random from the
distribution of delay times found in the \swift\ follow-up of IceCube
neutrino triggers \citep{Evans15}, which is a reasonably analogous ToO
programme. This is well modelled as a Lorentz distribution:

\begin{equation}
P(\tau_d) = \frac{N_l}{1+\left( 2\left[\frac{\tau_d -
C_l}{W_l}\right]\right)^2}
\label{eq:delay}
\end{equation}

\noindent where $P(\tau_d)$ is the probability of the delay being $\tau_d$, $N_l=1103, C_l=-1.95$ hr and $W_l=-8.97\tim{-2}$
hr. 

{\bf X-ray afterglow template}. For each simulation we selected at 
random one of the sGRBs listed in \cite{Rowlinson13}\footnote{i.e.\ with a
BAT $T_{90}\le 2$ s, and to which \swift\ slewed promptly upon detection.
}, for which a valid 
light curve fit exists in the live XRT GRB 
catalogue\footnote{http://www.swift.ac.uk/xrt\_live\_cat} 
\citep{Evans09}. We then took the automatic multiply-broken power-law 
fit to XRT light curve from this catalogue, and `shifted' it to the 
redshift of the GW simulation by applying luminosity distance and time 
dilation corrections. For those sGRBs where the redshift was not known, 
we assumed the mean value of 0.72 \citep{Rowlinson13}\footnote{\protect\cite{Davanzo14} used a slightly different
criteria for selecting sGRBs, and found a mean redshift of 0.85.}. We did not apply 
a k-correction, to keep the simulations as simple as possible, having 
first verified that this made little difference: even in the most 
extreme case (taking a short GRB from $z=2.6$ to simulate a NS-NS merger 
at 9 Mpc) the k-correction would make $\sqiglt10\%$ difference to the 
light curve normalisation. We can only use as templates those sGRBs
for which we have a valid light curve fit, which introduces a bias:
we have light curve fits for 29 of the sGRBs listed in 
\cite{Rowlinson13} (of the remaining 14, 6 were not detected by XRT, and for
the others we have only a single light curve bin, or a bin and an upper limit,
so no fit is available). Thus, we are really simulating the brightest (in terms 
of flux, not luminosity) 67\%\ of sGRBs observed by XRT. As will be seen in Section~\ref{sec:howbright},
this bias is not important when compared to other considerations.

In order to marginalise better over these different parameters, we 
performed this process (from deciding whether to shift the GW 
simulation onwards) ten times for each GW simulation, giving 2,350 
simulations for 2015, and 1,930 for 2016.

\subsection{Defining the XRT follow-up}

At the present time it is not known how many fields it is practical to 
observe with \swift\ immediately after a GW trigger\footnote{Investigative work is under way.}; nor do we know how 
many we would choose to observe, since a point of diminishing returns 
is likely to exist (i.e.\ where observing extra fields barely increases the probability of 
detecting the GRB). The number of fields observed is therefore also a parameter we wish to explore. Initially, we created up to 2,400 
XRT fields per GW simulation -- more than we expect to observe in 
reality --  by splitting the sky into a pattern of overlapping XRT 
fields, integrating the probability in the GW map within each field, 
and selected those with the highest probability (up to 2,400 fields, or 
where the probability of the event lying in a field fell below 
$10^{-6}$). We did this in two ways: (1) integrating the probability 
from the original GW probability map, and (2) by integrating the 
probability from the GW map convolved with the GWGC mask 
(Section~\ref{sec:galcat}). 

In order to realistically simulate XRT observations, the following information
was produced for each field, in addition to the probability of it containing the GRB.

\begin{figure}
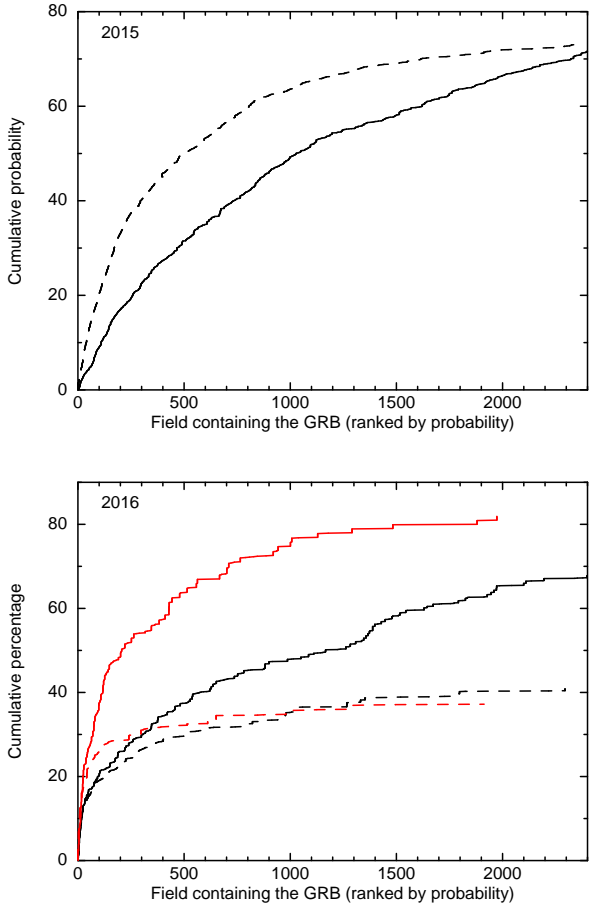

\begin{center}
\psfig{file=2015_whichField.eps,height=8.1cm,angle=-90}
\psfig{file=both_2016_whichField.eps,height=8.1cm,angle=-90}
\end{center}
\caption{The impact of targeting galaxies. When 2,400 XRT fields are created to cover the most probable region
of the ALIGO error region, the solid line shows the cumulative distribution of which field (ordered by probability)
the GRB actually lies in. The dashed line shows the same, if the ALIGO error region is first convolved with the GWGC
(Section~\ref{sec:galcat}). \emph{Top:} The 2015 ALIGO configuration (based on the {\sc bayestar}
probability map). Using the GWGC fewer fields are needed to reach a given probability of observing the GRB.
\emph{Bottom:} The 2016 ALIGO-AVirgo configuration:
the {\sc bayestar}-based results are in black, those derived from the full MCMC error region are in red.
Using the GWGC reduces the number of GRBs observed, because most of them occur beyond the 100 Mpc
limit of GWGC catalogue.}
\label{fig:whichField}
\end{figure}

\begin{enumerate}

\item{{\bf The mean X-ray background level}. This was drawn at random from the 
distribution of such levels seen in the 1SXPS catalogue\footnote{We
used the 0.3--10 keV background count-rate (`FIELDBG0') from all 1SXPS fields with a `FIELDFLAG' of
zero (i.e. they contained no artefacts or extended sources).} \citep{Evans14}.}

\item{{\bf Details of any catalogued X-ray sources in the field}. These were taken
from the RASS. The expected 0.3--10 keV XRT count rate was derived from the 
\emph{Rosat} count rate using {\sc pimms} and assuming a typical AGN 
spectrum: an absorbed power-law with $N_{\rm H}=3\tim{20}$ \cms, 
$\Gamma=1.7$. \label{stem:rass})}

\item{{\bf The 3-$\sigma$ RASS upper limit in the field}. This was derived using the RASS images, 
background maps and exposure maps\footnote{Obtained from
ftp://ftp.xray.mpe.mpg.de/ftp/rosat/archive}, and the Bayesian method
of \cite{Kraft91}. This was converted to an XRT count-rate using {\sc pimms} as above.}

\item{{\bf Details of uncatalogued X-ray sources coincidentally in the field}. 
We derived these using the  log $N$-log $S$ distribution of \cite{Mateos08}, which gives the sky density of 
extra-galactic X-ray sources as a function of flux. From this we determined
the expected number of sources unrelated to the GW trigger ($N_s$) in the 
field of view, with a flux of at least $10^{-13}$ erg \cms\ s$^{-1}$ 
(this limit was chosen as it is well below the detection limit of 
\swift-XRT in the exposure durations we are going to explore). Then, 
drew the number of sources ($N_{s,r}$) in this simulated field at random
from a Poisson distribution with a mean of $N_s$. If this number was
greater than the number of sources already identified in
step~\ref{stem:rass} then we assigned extra sources to the field, until it
contained ($N_{s,r}$) sources. The fluxes of these sources were drawn at random from the \cite{Mateos08}
log $N$-log $S$ distribution, up to a maximum flux equivalent to the RASS upper
limit for the field. We converted from flux to an XRT count-rate using the AGN spectrum
defined in step~\ref{stem:rass}.}

\item{{\bf The visibility windows of the field for \swift-XRT}, for an interval 
of 15 days starting at the trigger.}

\end{enumerate}

Having thus generated 2,400 fields, the most probable $N$ can be selected as inputs to
our simulated observations, where $N$ is one of the parameters the simulations are designed to explore.

\subsection{Simulating the observations}
\label{sec:obssim}

Having defined the fields for XRT follow-up, we simulated the 
actual XRT observations for different observing strategies, to evaluate the
impact of those strategies on how many sGRBs which trigger ALIGO/AVirgo
we can identify with \swift-XRT. The strategies were based on two parameters:
the number of XRT fields ($N$), and the exposure per observation ($e$). 
We explored values of $N=50, 100, 200, 600, 1200$ and $1800$, and $e=10,50,100$ and
500 s. This was done both for the fields based on the raw GW
probability map and those based on the convolution with the GWGC
catalogue.

We simulated 15 days of observations, recording after each field was observed whether the
GRB had been identified, and also how many uncatalogued sources
which were not the GRB had been identified. Full details of how
the fields to be observed were selected are given in Appendix~\ref{app:sims}.

\begin{figure}
\begin{center}
\psfig{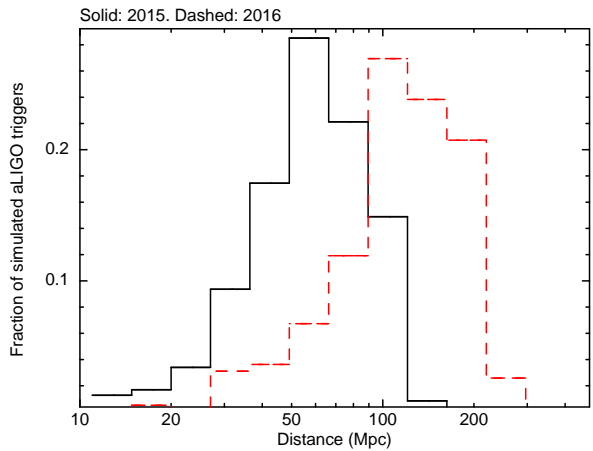}
\end{center}
\caption{The distribution of the simulated NS-NS mergers detected by the simulated 2015 ALIGO 
configuration (solid black) and 2016 ALIGO+AVirgo configuration (dashed red). Both plots are normalised
to 1.}
\label{fig:distance}
\end{figure}

\label{sec:galtarg}
\begin{figure*}
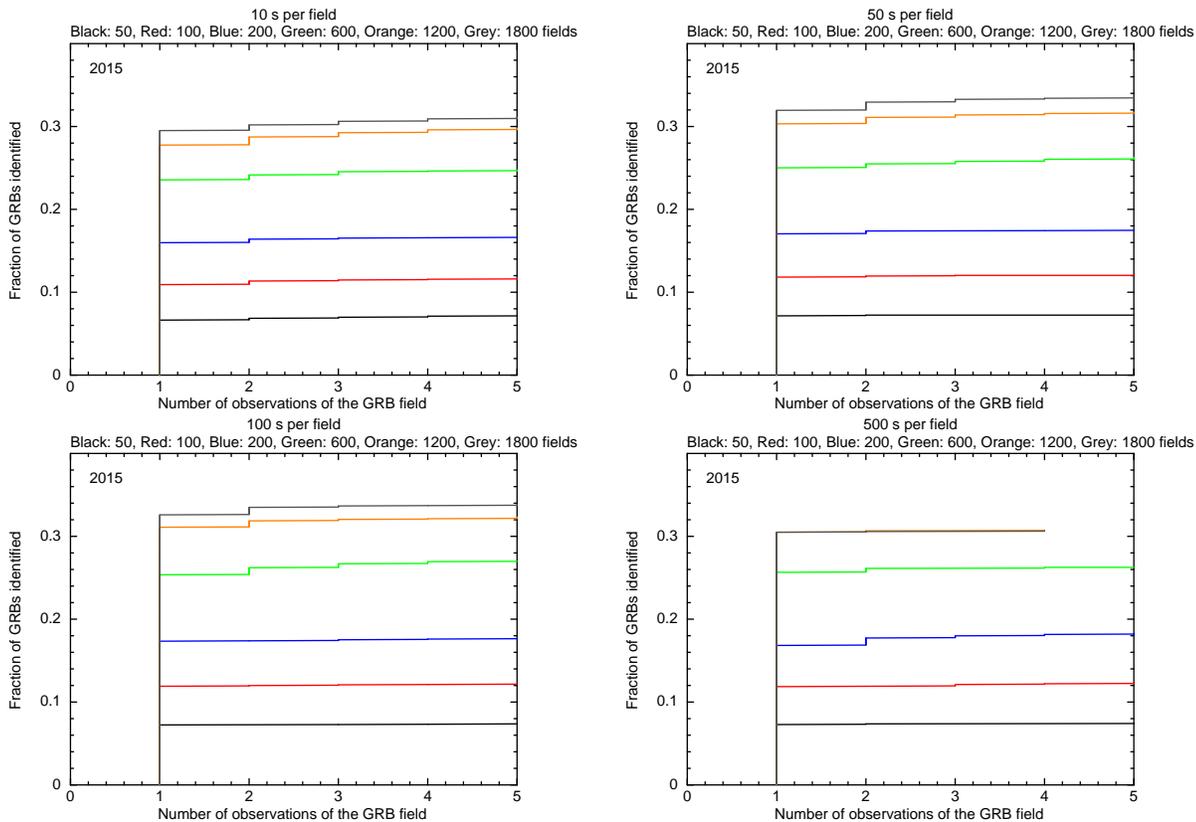

\begin{center}
\hbox{
\psfig{file=nObsToID_2015gal_10.eps,height=8.1cm,angle=-90}
\psfig{file=nObsToID_2015gal_50.eps,height=8.1cm,angle=-90}
}
\hbox{
\psfig{file=nObsToID_2015gal_100.eps,height=8.1cm,angle=-90}
\psfig{file=nObsToID_2015gal_500.eps,height=8.1cm,angle=-90}
}
\end{center}
\caption{The fraction of GRBs identified as a function of how many observations of all fields
have been performed. This is for the 2015 configuration simulations, with galaxy targeting.
The 4 panels are for the 4 values of $e$ (exposure per observation)
we simulated, and the different colours refer to different values of $N$ (the number of fields).}
\label{fig:nObsToID}
\end{figure*}

\section{Result 1: Should we target galaxies?}
\subsection{2015}

To investigate the effect of convolving the GW error region with the 
GWGC (Section~\ref{sec:galcat}), we recorded which field (ordered 
by decreasing probability) the GRB actually lay in. Not all GRBs lay within
the 2,400 most probable fields, regardless of whether galaxy targeting was employed.
For the 2015 ALIGO configuration, the number of bursts which did lie in these regions was barely affected
by use of the GWGC: 1,291 (/2,350) GRBs lay inside the most-probable 2,400 fields
whether or not the GWGC was used; 425 GRB only lay in these fields if the GWGC was used, and 394
were only present if the GWGC was not used. However, use of the GWGC enables the GRB to be observed quicker.
Fig.~\ref{fig:whichField} shows the cumulative distribution of which field contained 
the GRB, for both approaches. Convolving the ALIGO error
with the GWGC means that we need observe far fewer fields to image the GRB. For example, to contain
50\%\ of the ALIGO triggers if we use GWGC convolution, we need observe `only' 500 fields; without
the convolution 1,030 fields are needed.

In summary: for 2015 it is best to target galaxies provided the GW events arise
from NS-NS binaries. For NS-BH binaries, the horizon distance of aLIGO is greater,
and the situation is more akin to the 2016 NS-NS case discussed below.

\subsection{2016}
For 2016 the situation is very different. Unlike 2015, most of the NS-NS mergers detected
were more than 100 Mpc away (see Fig.~\ref{fig:distance}) and so do not reside in GWGC galaxies.
In consequence, targeting those galaxies reduces the number of NS-NS mergers that lie in
XRT fields.
Thus for the 2016 configuration, targeting GWGC galaxies is detrimental to our chances of detecting
the X-ray counterpart to a GW trigger. Without galaxy targeting, the typical number of fields required to 
image the merger location in 2016 is similar to that in 2015. That is,
the 2016 configuration increases the horizon distance without reducing the size of
the error region. However, this is a software issue. \cite{Singer14} show (their fig.~3) 
that for the 2016 configuration the fast {\sc bayestar} algorithm which 
is used to construct the GW error region does not perform particularly 
well, and results in a larger-than-necessary error region. The full MCMC 
approach typically gives much smaller error regions; 
however it also takes much longer to run, (`hours to days' according to 
\citealt{Singer14}), which makes waiting for it impractical. 
\cite{Singer14} were optimistic that, by 2016, the time delay may 
have been dramatically reduced. Indeed, the latest news (Singer, private 
communication) is that the MCMC process has been made faster, and 
{\sc bayestar} has been improved and now performs as well as 
the full MCMC for the 2016 configuration. Therefore, the full MCMC
probability maps from \cite{Singer14} are more representative of what the
rapidly available {\sc bayestar} maps will actually be like in 2016, 
so we repeated our simulations for 2016, this time using the MCMC probability 
map instead of {\sc bayestar}. Unfortunately, only 100 of the 193 
on-axis mergers for 2016 in \cite{Singer14} had MCMC probability maps, 
resulting in lower statistics. However, Fig.~\ref{fig:whichField} shows 
that, with MCMC, the situation is dramatically improved. It is of course 
still true that use of the GWGC is detrimental, since most mergers in 
2016 lie beyond that catalogue's distance horizon, however even without 
galaxy targeting, the smaller error regions available in 2016 actually 
mean that it takes fewer fields on average to observe the location of 
the GRB than with galaxy targeting in 2015. As noted earlier 
(Section~\ref{sec:galcat}), other catalogues such as 2MPZ may be 
appropriate for use in 2016. 

In summary: for 2016 the GW regions should not be convolved with the GWGC. Other
galaxy catalogues may be appropriate, provided their completeness out to 200--300 Mpc
can be derived and shown to be substantial.

\section{Result 2: Optimal observing strategy}
\label{sec:strategy}

\begin{figure}
\begin{center}
\psfig{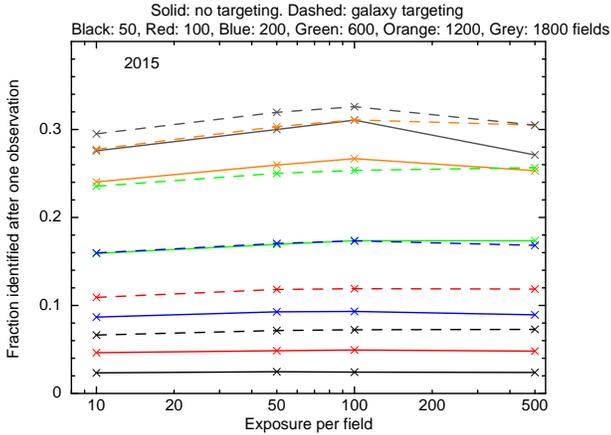}
\end{center}
\caption{The fraction of GRBs identified in the 2015 simulations, after a single observation of
all $N$ fields, as a function of the exposure per observation. The solid lines show the result
when the $N$ fields are based on the raw ALIGO probability map, and the dashed lines for if the map is
first convolved with the GWGC.}
\label{fig:nObs_effect}
\end{figure}

\subsection{2015}
\label{sec:res2015}

The previous section demonstrated that, for the 2015 configuration,
it is preferable to target galaxies in the ALIGO error regions.
In Fig.~\ref{fig:nObsToID} we show the fraction of the galaxy-targeting simulations for which the GRB was identified,
as a function of how many observations of the $N$ fields were conducted. 
This is shown for all values of $N$ (different coloured lines) and $e$ (different panels)
used in our simulations. Recall that we only simulated on-axis GRBs, therefore this
is the fraction of such GRBs, not the fraction of merger events in total. For all of the configurations we tested, 
almost every GRB identifiable is identified in the first observation of its field:
repeat observations have little value, as far as identifying the bright X-ray object is concerned.
The exposure dependence is illustrated more clearly in Fig.~\ref{fig:nObs_effect}, where we show
the fraction of GRBs identified after a single observation of all $N$ fields, as a function
of the exposure time per observation. Little exposure dependence is seen, except for the combination of
large values of $e$ and $N$, where the fraction identified drops slightly. This is because, in such cases
it takes so long to reach the field containing the GRB that fades below the RASS limit before it is observed.
Fig.~\ref{fig:nObs_effect} also includes the simulations with and without
galaxy targeting, clearly demonstrating the value of galaxy targeting in 2015.

\begin{figure}
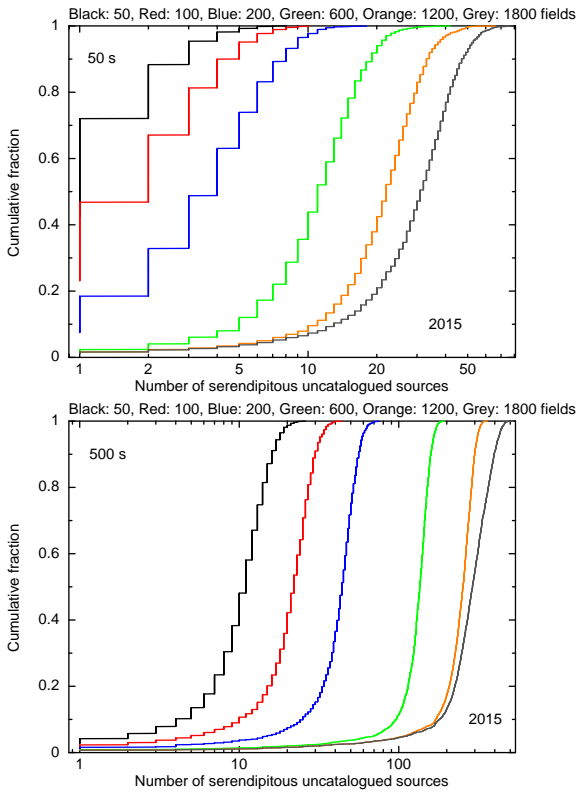

\begin{center}
\psfig{file=nseren1_gal50s_2015.eps,height=8.1cm,angle=-90}
\psfig{file=nseren1_gal500s_2015.eps,height=8.1cm,angle=-90}

\end{center}
\caption{The cumulative distribution of the number of uncatalogued sources unrelated to the GW trigger detected 
in the 2015 simulations (with galaxy targeting), 
assuming observations of 50 s (top panel) or 500 s (bottom). If no bright source is detected in the observations,
all of these uncatalogued sources would have to be reobserved to search for signs of fading.}
\label{fig:nseren}
\end{figure}

In a very small number of the simulations (\til0.05\%) the GRB was detected,
but was not bright enough to be identified. In this case, a second observation could be performed to identify
fading behaviour in the source. However, this is a very inefficient approach. As Fig.~\ref{fig:nseren} shows,
most strategies will result in the detection of new, uncatalogued sources which are not the GRB, all of which would
have to be followed up; and as already noted, the number of cases where the GRB is among these sources
is extremely small.

\subsubsection{False positives}
\label{sec:falsebright}

\begin{figure}
\begin{center}
\psfig{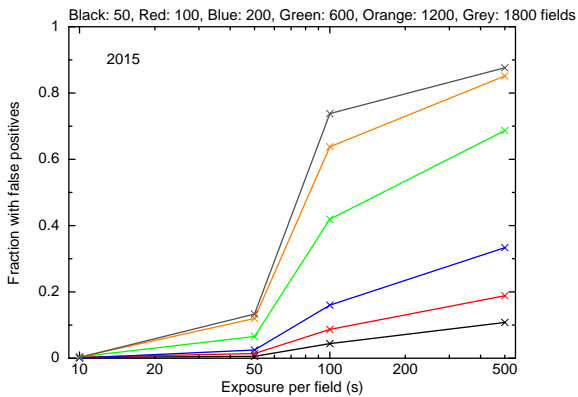}
\end{center}
\caption{The fraction of simulations in which a source unrelated to the GW trigger was classified as a potential afterglow,
because its measured flux was above the RASS limit due to Poisson noise. The data shown here are for the
2015 simulations, with galaxy targeting.}
\label{fig:serenBright}
\end{figure}

We noted earlier (Section~\ref{sec:falsePos}) that the expected rate of X-ray
transient sources is 1 per 1.64 Ms of exposure time. So for the case of 600 fields, observed
for 50 s each, there is only a 2\%\ chance of serendipitously observing a X-ray transient unrelated to the GW trigger.
There is, however, an additional source of potential contamination: due to Poisson noise,
an uncatalogued source with a flux below the RASS limit will sometimes be measured to have a flux 
more than 1-$\sigma$ above this limit, thus meeting one of our criteria to be identified as the 
likely GRB afterglow. Our simulations included Poisson noise, and we recorded how many of the unrelated
sources detected would be thus classified as a possible afterglow.
Fig.~\ref{fig:serenBright} shows the fraction of simulations in which this happened for at least one unrelated source.
This increases sharply for exposures longer than 50 s.
This means that we cannot interpret the discovery of a bright ($>$RASS limit), uncatalogued source to be an unambiguous
detection of the sGRB: it could simply be a fainter source subject to Poisson errors. Also, there could be multiple `bright' sources
detected in the follow-up of a single GW trigger. Therefore, it will be necessary to re-observe any such sources for \til1 ks each
to confirm which (if any) of them was a transient event. The maximum number of such false positives detected in
any of our simulations is 19, so this confirmation observation requires only  a small
investment of exposure time, compared to that used to detect the sources.

\subsection{2016}
\label{sec:2016}

\begin{figure}
\begin{center}
\psfig{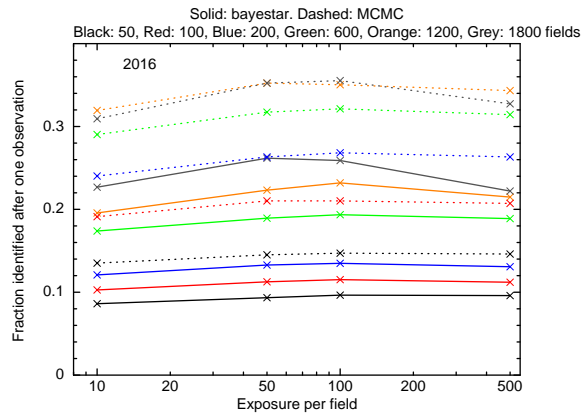}
\end{center}
\caption{The fraction of GRBs identified in the 2016 simulations, after a single observation of
all $N$ fields, as a function of the exposure per observation. The solid lines show the result
if the {\sc bayestar} error map is used, and the dashed lines are for
the full MCMC error map.}
\label{fig:nObs_effect_2016}
\end{figure}

We performed two sets of simulations for 2016, one based on the {\sc bayestar} error maps
and one based on the full MCMC maps. Since Fig.~\ref{fig:whichField} showed that targeting GWGC 
galaxies was detrimental in 2016, we only simulated observations based on the raw GW error maps.
As with the 2015 simulations almost all identifiable GRBs were identified
in a single observation. Fig.~\ref{fig:nObs_effect_2016} shows the fraction of GRBs detected as a function of
exposure time, for the different simulations, an analogue of Fig.~\ref{fig:nObs_effect}. Using the {\sc bayestar} maps
we identified a lower fraction of GRBs than in the 2015 simulations, although for the $N\le200$ simulations, 
the difference is less obvious: the fraction identified in 2016 is comparable to the 2015 follow-up 
with no galaxy targeting. In contrast, using the full MCMC-based error map in 2016 we identified the GRB in a
larger fraction of simulations than in 2015.

One curious feature of  Fig.~\ref{fig:nObs_effect_2016} is that, for the $N=1,800$ simulations
based on the full exposure map, fewer GRBs were identified than the $N=1,200$ simulations at $e=500$ s.
This is an artefact of the way the simulations decide in which order to observe the XRT fields (Appendix~\ref{app:sims}), which
means that while the fields in the $N=1,200$ simulation are the most probable 1,200 fields in the $N=1,800$ simulation, they are not
necessarily the first 1,200 to be observed in the latter simulations. This, combined with the spatial distribution of probability
in the 2016 MCMC error maps, which tends to contain distinct `clumps', means that on average the field which
actually contained the GRB was observed later in the $N=1,800$ simulations than in the $N=1,200$ case. As a result,
the GRB was more likely to have faded to below the detection threshold or RASS limit.

\begin{figure}
\begin{center}
\psfig{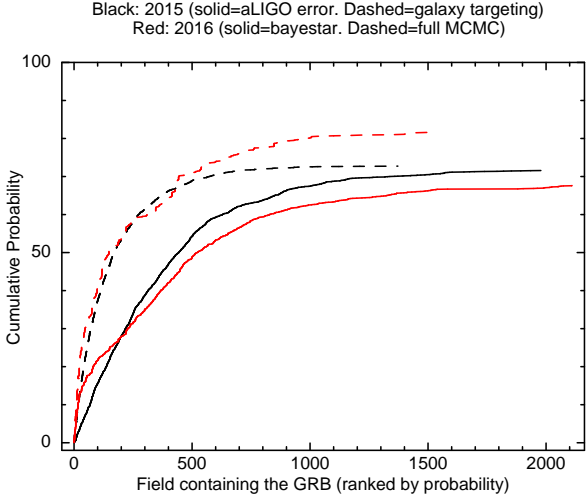}
\end{center}
\caption{As Fig.~\ref{fig:whichField} but showing the effect of using information from a \emph{Fermi}-GBM trigger
simultaneous with the GW trigger. The black lines are for the 2015 ALIGO-detected mergers, with the dashed
and solid lines showing respectively whether the GWGC is or is not used. The red lines show the 2016 ALIGO/AVirgo 
objects, with the solid line showing the {\sc bayestar} probability maps and the dashed line the full MCMC map.}
\label{fig:gbm}
\end{figure}

\section{If \emph{Fermi} triggers as well}
\label{sec:GBM}

If \swift-BAT and ALIGO/AVirgo trigger on the same event, the discussion 
in this paper is moot: the BAT error region fits comfortably in the XRT 
(and UVOT) field of view, and the satellite will perform the standard automatic
follow-up process which has resulted in the rapid detection of the X-ray 
afterglow of almost all sGRBs detected by \swift-BAT to 
date. However, as noted earlier, this is unlikely. A more 
probable scenario for the simultaneous GW and EM detection of the same 
event is for \emph{Fermi\/} to trigger on the GRB. \emph{Fermi\/} 
contains two instruments: the Gamma-ray Burst Monitor (GBM; 
\citealt{Meegan09}) and Large Area Telescope (LAT; \citealt{Atwood09}). GBM is sensitive 
to triggers over the entire sky which is unocculted by the Earth, 
although has poor positional accuracy, with typical uncertainties 
of many degrees. LAT has a smaller field of view (2.4 sr), and
if GBM detects a bright GRB \emph{Fermi\/} with a sufficiently high peak flux, \emph{Fermi\/}
automatically slews to move the GRB into the LAT field of view \citep{Ackermann13,vonKienlin14}. LAT
positions can be much better than those from GBM,  down to \til0.1\deg.
While it is beyond the scope of this work to simulate the probabilities of
these instruments triggering on GW-detected sGRBs, we consider how such triggers
will influence \swift\ observations. Note that, due to their proximity,
we would expect such sGRBs to be bright enough to trigger \emph{Fermi\/} (or indeed BAT or the Interplanetary Network [IPN])
unless they are \til100 times fainter than the typical \swift-detected sGRBs (see Section~\ref{sec:howbright}).

\subsection{If LAT triggers}

While the LAT error region is highly variable, at the lower end they are small enough to fit inside a single
XRT field of view. For less-well constrained positions, \swift\ has often 
observed the LAT error region in a small number of tiles (typically 4, sometimes 7; e.g.\ \citealt{Evans13GCN}).
This allows us to rapidly cover the entire error region,
and build up long exposures of multiple kiloseconds per tile in much less time that it takes
to observe each field once if we have only a GW trigger. These longer exposures allow
easy identification not just of a bright source, but also of a faint, but clearly fading object.
Due to the small area covered, the possible presence of unrelated sources is
not a significant issue, since the burden of following multiple sources to confirm their
behaviour is much reduced when those sources are close together. A drawback to LAT triggers
is that nearly all of them are found in ground processing, which means that are not detected
until many hours after the event. However, a nearby burst will likely also be bright, increasing the
likelihood of an onboard LAT trigger. Further, even if \swift\ has already started tiling
the GW error region, should a LAT error region become available, we could immediately
terminate the ongoing observations and point \swift\ at the LAT region.

\subsection{If GBM triggers}

Even if LAT does not detect the GRB, a GBM detection simultaneous with 
an ALIGO/AVirgo trigger has the potential to significantly reduce the 
sky area we need to search. The median 1-$\sigma$ statistical error radius for a 
GBM-detected sGRB\footnote{Taken from the Fermi GRB catalogue, 
\protect\cite{Gruber14,vonKienlin14}, 
http://heasarc.gsfc.nasa.gov/W3Browse/fermi/fermigbrst.html, selecting 
objects with $T_{90}<2$ s and {\sc flux\_batse\_64}$>2.37$ s$^{-1}$, 
following \protect\cite{Wanderman15}.} is 7\deg. A systematic error is 
also present \citep{Connaughton15}, which we simplistically treat as a 
Gaussian with $\sigma$=3.7\deg, giving an overall error median error of 
8\deg. For each simulated GW event in our samples, we selected at random 
the location of the centre of the GBM error region, assuming a circular 
region, with a Gaussian radius, with $\sigma_{\rm tot}=8\deg$, and then created 
the list of XRT fields to observe as in Section~\ref{sec:simulate}, but 
also only accepting fields that lay within 3-$\sigma_{\rm tot}$ of the
simulated GBM position. We then determined how many fields we would have 
to observe (to a maximum of 2,400) in order to observe the GRB location. 
The result is shown in Fig.~\ref{fig:gbm}, which can be compared with 
Fig.~\ref{fig:whichField}. This shows, not surprisingly, that if a GBM 
localisation is available, it can reduce the number of XRT fields we need 
to observe significantly, e.g.\ in 2015, with galaxy targeting, the 
number of XRT fields needed to observe 50\%\ of the GRBs is just 190 
(compared to 500 without GBM: see Section~\ref{sec:galtarg}).

\section{Are detectable, nearby sGRBs likely?}
\label{sec:howbright}

\begin{figure*}
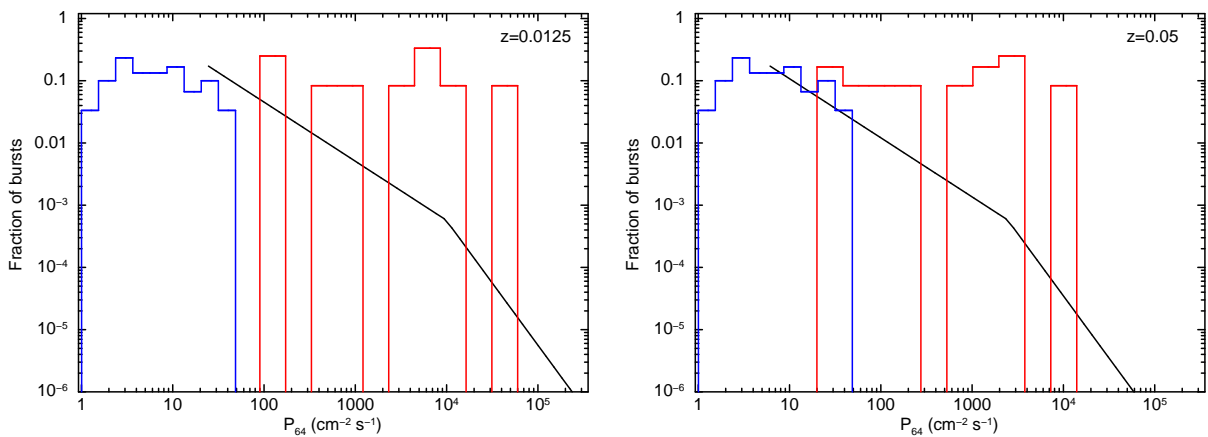

\begin{center}
\hbox{
\psfig{file=Swift_109Mpc.eps,height=8.1cm,angle=-90}
\psfig{file=Swift_222Mpc.eps,height=8.1cm,angle=-90}
}

\end{center}
\caption{A comparison of the observed brightness of the \swift-sGRBs without redshift,
to what would be expected from an sGRB within the GW horizon distance. $P_{64}$ is the
peak 64-ms photon flux. The blue histogram shows the distribution of $P_{64}$ for the Swift short GRBs
without redshift. The red histogram shows the distribution of short GRBs with redshift measurements, with the peak flux corrected to a
distance of 100 (left panel) or 200 (right panel) Mpc. The resultant peak fluxes are orders
of magnitude larger than anything Swift has observed to date. The solid black line
shows the theoretical sGRB luminosity function of \protect\cite{Wanderman15}, converted
to $P_{64}$ at the two distances.}
\label{fig:howBright_swift}
\end{figure*}

So far we have focussed on how best to search for an XRT counterpart to a GW trigger,
but another important consideration is whether a GW trigger from a nearby, on-axis sGRB is likely.
Various works have attempted to calculate the rate of nearby sGRBs, either from the theoretically expected merger
rate \citep[e.g.][]{Abadie10} or the rate of observed sGRBs \citep[e.g.][]{Coward12}. As noted
earlier, these estimates have large uncertainties, for example, \cite{Coward12} give an
sGRB rate density of \til8 -- 1100 Gpc$^{-3}$ yr$^{-1}$. \cite{Metzger12} fitted the observed
redshift distribution of \swift-detected sGRBs with known redshift, divided this by the fraction
of sGRBs with known redshifts, and from that inferred that ``a few \emph{Fermi} bursts over the past few years
probably occurred within the ALIGO/Virgo volume.'' However these approaches do not consider one important
observational constraint: unless the nearby sGRBs are systematically less luminous than the 
ones for which we have redshifts, they should appear much brighter than those -- orders of magnitude brighter.
Is there evidence for any such bright objects among the sGRBs detected to date?

\begin{figure}
\begin{center}
\psfig{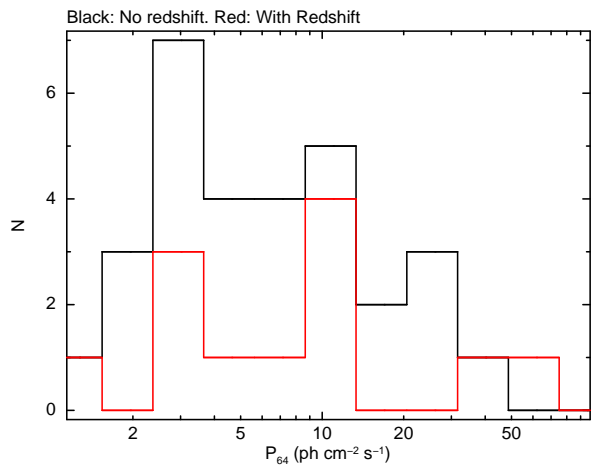}
\end{center}
\caption{The distribution of observed peak 64-ms photon flux from the
Swift GRBs with redshift (red) and without (black). The K-S test gives a 0.43 probability
that the two samples come from the same parent distribution.}
\label{fig:samples}
\end{figure}

To investigate this we want to compare the expected brightness of an sGRB inside the ALIGO/Virgo 
horizon with the distribution of observed sGRB brightnesses. We define the brightness as the 64-ms peak photon
flux ($P_{64}$) from the prompt emission, following \cite{Wanderman15}. For the \swift-detected sGRBs with known redshift listed in 
\cite{Wanderman15}, we derived the 15--150 keV $P_{64}$ values that would have been measured had the GRB been at
$z=0.025$ (\til100 Mpc) and $z=0.05$ (\til200 Mpc); these are shown as red histograms in Fig.~\ref{fig:howBright_swift}.
This sample  is likely biased towards intrinsically brighter objects (i.e.\ 
objects which are easier to detect), so we have also converted the theoretical luminosity
function of short GRBs from \cite{Wanderman15} into the distribution of 15--150 keV $P_{64}$ at these two redshifts;
this is shown as the black line in Fig.~\ref{fig:howBright_swift}. For comparison, we also show the distribution of
the observed 15--150 keV $P_{64}$ values for the \swift-BAT detected sGRBs without a known redshift (blue histograms).
The $P_{64}$ distribution of the \swift\ sGRBs with no redshift is clearly not consistent with them being
identical to the known-redshift sample but lying at low redshift: in this case the blue histograms would
appear similar to (or at least share significant overlap with) the red ones.
It is possible that a small number of the no-redshift sample were at \til200 Mpc,
however they must have been of lower intrinsic luminosity than the
with-redshift sample, since they just barely overlap with what the latter sample would
have looked like at 200 Mpc. We certainly cannot rule out that some 
of the no-redshift sample lay at 100--200 Mpc and came from the faintest end of
the \cite{Wanderman15} luminosity function, although this faint end is not well constrained.
There are, however, two objections to this premise. First, the \emph{observed} $P_{64}$ distributions
of the with- and without-redshift \swift\ sGRB samples are very similar (Fig.~\ref{fig:samples}):
a Kolmogorov-Smirnov test gives the probability that the two samples are drawn from the
same distribution as 0.43; so if the no-redshift sample does indeed contain bursts that are of lower
luminosity than the with-redshift sample, the redshift distributions of the two samples have fortuitously
been tuned correctly to give apparently identical $P_{64}$ distributions, which seems a little far-fetched.
Secondly, if any of the no-redshift sample really were within 200 Mpc, we would expect
the host galaxy to have been found in optical follow-up, and hence the GRB to have a redshift.
\cite{Tunnicliffe14} explored this question and found tentative evidence that one or two
`hostless' sGRBs could have arisen from stars kicked out of nearby ($<200$ Mpc) galaxies
but that these would have surprisingly low jet energies. For example, they found a galaxy 
at a distance of 80 Mpc, which \emph{could} be the host of GRB 111020A,
but this would require the isotropic equivalent for that GRB to be just $10^{46}$ erg, several 
orders of magnitude lower than typical sGRBs. These considerations suggest that no compelling evidence 
exists for a BAT-detected sGRB within 200 Mpc.

\begin{figure*}
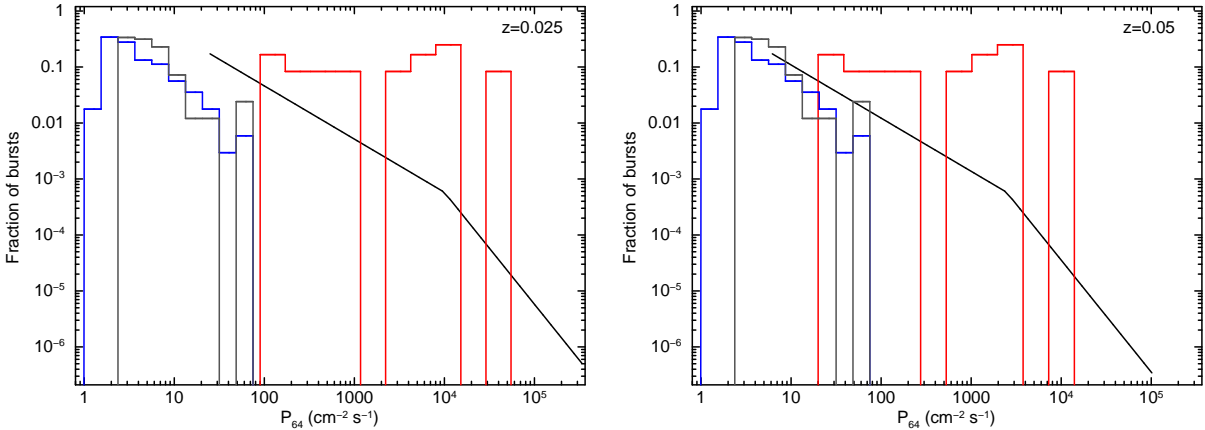

\begin{center}
\hbox{
\psfig{file=BATSE_109Mpc.eps,height=8.1cm,angle=-90}
\psfig{file=BATSE_222Mpc.eps,height=8.1cm,angle=-90}
}
\end{center}
\caption{As Fig.~\ref{fig:howBright_swift} but for the 50--300 keV band. The blue and grey histograms
show the distribution of observed peak 64-ms photon flux for the BATSE and GBM sGRBs respectively -- these have no measured
redshift.}
\label{fig:howBright_batse}
\end{figure*}

\swift-BAT detects fewer sGRBs than GBM does or BATSE did, so we have 
repeated the above test, for the BATSE and GBM bursts. The results are shown in Fig.~\ref{fig:howBright_batse}.
For these, we derived $P_{64}$ in the 50--350 keV band used by those instruments. We again plotted the 
expected $P_{64}$ values for the with-redshift \swift\ sGRBs (corrected to the 50--350 keV band), had they been at 100 or 200 Mpc (red histograms).
The BATSE and GBM observed $P_{64}$ distributions are shown in the blue and grey histograms respectively.
As with the \swift\ data, we cannot rule out the possibility that a few of the 
sGRBs detected by BATSE or GBM could have been within 100--200 Mpc from Earth, however,
again, these must have been significantly fainter (typically at least an order of magnitude) than the \swift-detected GRBs with redshift.
A substantial fraction of the observed GBM and BATSE bursts are consistent with being at 200 Mpc, provided
they come from the low-luminosity end of the \cite{Wanderman15} luminosity function. However, this cannot be systematically
the case, since the GBM and BATSE samples must contain the more luminous GRBs that \swift\ has detected as well.

Two historic events warrant explicit mention: GRBs 051103 and 070201 were both very bright events, localised by the IPN,
and with large error regions that partially intersected M81 and M31 respectively. However, LIGO observations
showed that GRB 070201 could not have been a NS-NS or NS-BH merger in M31 \citep{Abbott08}, suggesting instead
that it was a giant flare from a soft gamma repeater (SGR), or a more distant sGRB coincidentally close in projection to M31.
The case for GRB 051103 is less clear, and an SGR origin of this event cannot be ruled out. If it were
a sGRB in M81 then it had a very low energy release (isotropic equivalent of $\til10^{46}$ erg) and an
extremely faint afterglow \citep{Hurley10}.

These considerations mean that we cannot conclusively rule out that some of the observed
GRBs without redshifts could have been nearby -- particularly those detected by BATSE, GBM or IPN, where the error regions are too
large for host galaxy searches. However, we can state that, were this so, these nearby bursts were also underluminous
compared to the with-redshift sample of \swift\ sGRBs. In more than 20 years of observations, no sGRB
has been detected consistent with being less than 200 Mpc away and having a prompt luminosity typical of the \swift\ sample.
This means that events such as those we have simulated in this work 
are rare. So if ALIGO/Virgo does detect a binary NS merger, it is likely to be significantly fainter (1--2 orders of magnitude)
than in our simulations, and therefore much harder to detect. More likely is that a NS-NS merger
detected by ALIGO/AVirgo will be off-axis, that is, the jet will not be pointed towards Earth. For such objects
no prompt gamma-ray emission is seen from Earth, so such objects are not in our observed GRB sample. Afterglow emission is still expected
from some such objects however, and we now briefly consider how detectable such emission may be.

\subsection{Off-axis GRBs}
\label{sec:offaxis}

\begin{figure*}
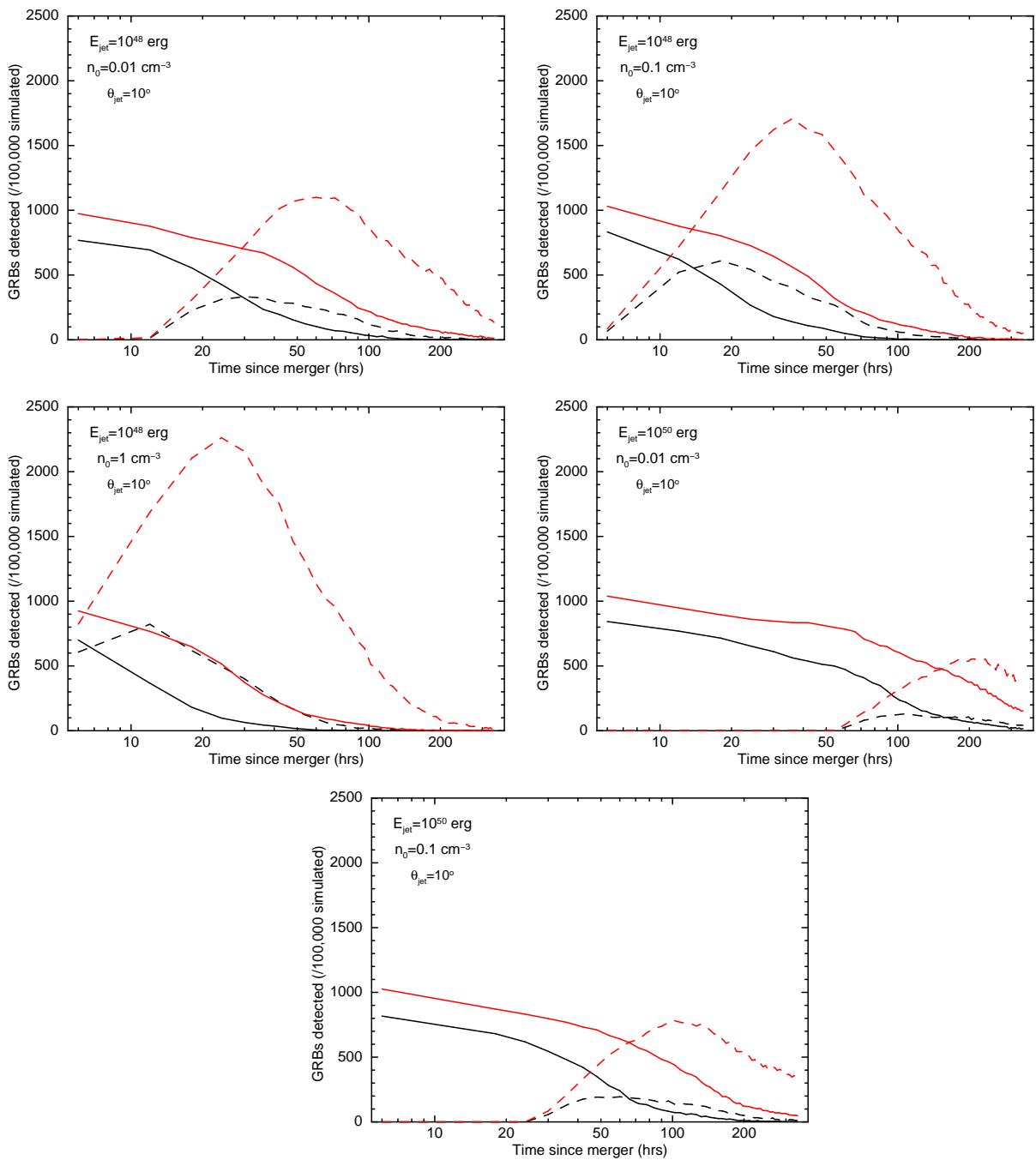

\begin{center}
\hbox{
\psfig{file=angleeffect_E48_n001.eps,height=8.1cm,angle=-90}
\psfig{file=angleeffect_E48_n01.eps,height=8.1cm,angle=-90}
}
\vspace{0.25cm}
\hbox{
\psfig{file=angleeffect_E48_n1.eps,height=8.1cm,angle=-90}
\psfig{file=angleeffect_E50_n001.eps,height=8.1cm,angle=-90}
}
\vspace{0.25cm}
\psfig{file=angleeffect_E50_n01.eps,height=8.1cm,angle=-90}
\end{center}
\caption{The number of bursts (out of 100,000 simulated) detectable in a 50 s (black) or 500 s (red)
observation, as a function of how long after the merger the observation begins. The solid lines are
bursts viewed `on-axis', that is the 10\deg\ jet is oriented towards the observer. The dashed lines are
the off-axis events. The different panels correspond to different values of jet energy ($E_{\rm jet}$) and circumburst
density ($n_0$). At early times the on-axis bursts dominate the visible population, despite comprising
only 1.5\%\ of all mergers. At late times, as the afterglow becomes visible off-axis, the
off-axis population starts to dominate.
}
\label{fig:offaxis}
\end{figure*}

Although there are no X-ray observational constraints on the properties of
GRBs viewed off-axis (by which we mean that the binary inclination $i$ is larger than the jet opening angle, $\theta_j$),
there have been theoretical predictions \citep[e.g.][]{vanEerten11,Ghirlanda15}.
In order to give some indication of the potential observability of off-axis sGRB afterglows with \swift\
we have taken a highly simplified approach. First, we assumed that the time at which the afterglow
becomes visible off-axis (\toa), for a binary at inclination $i$ radians, is the same as the time at which a jet-break
would occur if the jet opening angle ($\theta_j$) was $i$:

\begin{equation}
\toa = \left(\frac{i}{0.13}\right)^\frac{8}{3} \left(\frac{E_{52}}{n_0}\right)^\frac{1}{3}
\label{eq:toa}
\end{equation}

\noindent where $E_{52}$ is the energy in the jet ($E_{\rm jet}$) in units of $10^{52}$ erg, $n_0$ is the 
circumburst density in units of cm$^{-3}$, and \toa\ is in the rest-frame of the GRB.

\begin{figure*}
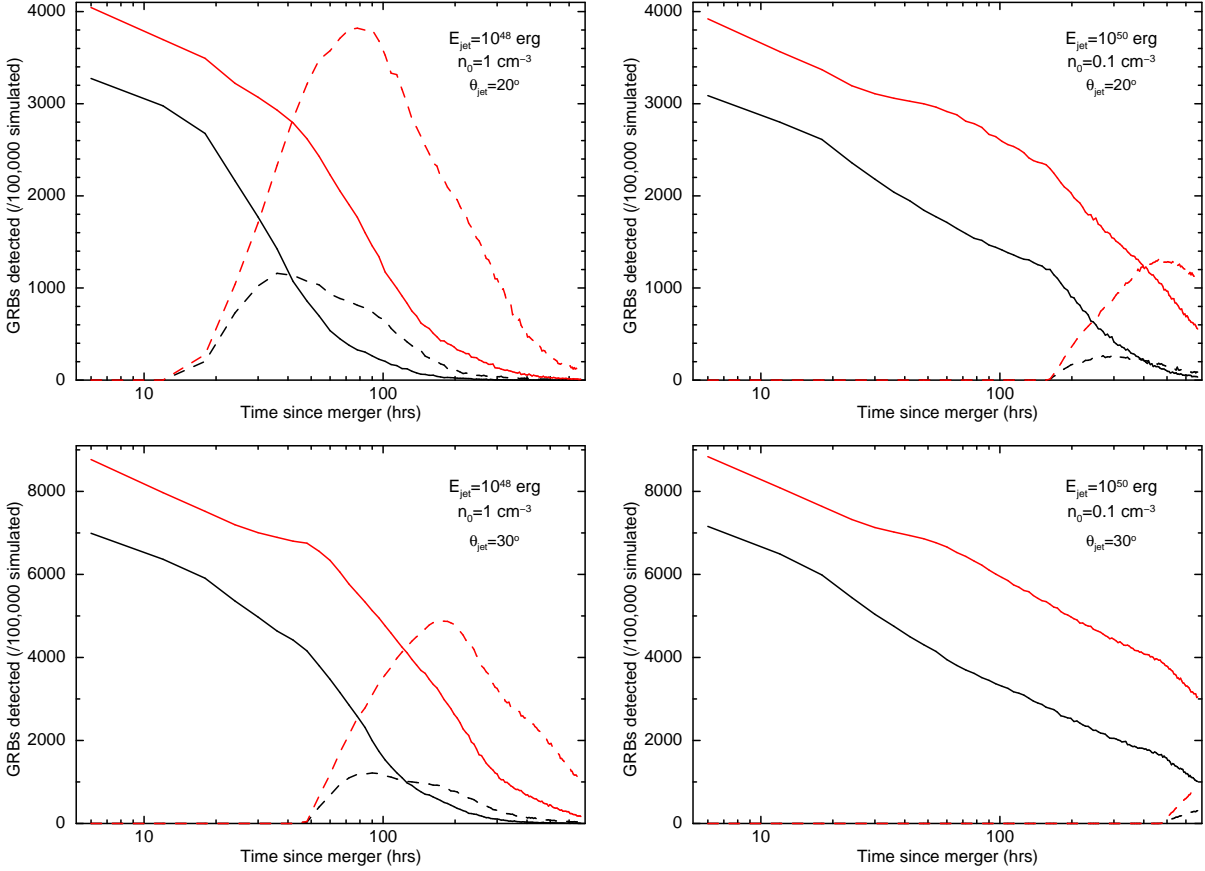

\begin{center}
\hbox{
\psfig{file=j20_angleeffect_E48_n1_j20.eps,height=8.1cm,angle=-90}
\psfig{file=j20_angleeffect_E50_n01_j20.eps,height=8.1cm,angle=-90}
}
\vspace{0.25cm}
\hbox{
\psfig{file=j30_angleeffect_E48_n1_j30.eps,height=8.1cm,angle=-90}
\psfig{file=j30_angleeffect_E50_n01_j30.eps,height=8.1cm,angle=-90}
}
\end{center}
\caption{How many on/off-axis afterglows are detected as a function of time for jet angles of
20\deg\ (top panels) and 30\deg (bottom panels). Rather than reproducing all the panels
of Fig.~\ref{fig:offaxis}, we show only the densities which give the most off-axis
bursts for each of the energies investigated. The solid lines are
bursts viewed `on-axis'. the dashed lines are
the off-axis events.
}
\label{fig:offaxis_wide}
\end{figure*}

Rather than the generous value of 30\deg\ used earlier, we began by assuming a jet 
opening angle of 10\deg\ for all bursts \citep[e.g.][]{Duffell15}. To simulate a GRB, we assigned 
it a distance $D$ up to 200 Mpc (with probability $P(D)\propto D^2$) and 
a binary inclination $i$ in the range 0--90\deg\ (with probability $P(i) \propto\sin i$). We selected 
an observed short GRB X-ray afterglow to use as a template, as in 
Section~\ref{sec:simulate}, but modified the light curve slightly to 
include a jet break (the jet break time calculated as in 
equation~\ref{eq:toa}, but replacing $i$ with $\theta_j=10\deg$), after 
which the light curve decayed with a power-law index of $-2.5$. For 
on-axis binaries ($i\le10\deg$) the brightness of the GRB afterglow at a given 
time was taken from this light curve. For off-axis binaries, we assumed 
there was no X-ray emission before \toa, after which the brightness 
followed the on-axis (post jet-break) light curve just defined. Comparison with the simulated light 
curves of 
\cite{vanEerten11}\footnote{
http://cosmo.nyu.edu/afterglowlibrary/sgrb2011.html} shows that this 
simplification is not unreasonable. We then calculated whether the GRB 
would be detected in observations of 50 s or 500 s, as a function of the 
time since the merger event (for these simple simulations we assumed the 
XRT background level was always $10^{-6}$ ct s$^{-1}$ pixel$^{-1}$, a typical value
from 1SXPS; \citealt{Evans14}).

We simulated 100,000 GRBs for each combination of $E_{52}$=0.01 and 
0.0001 (following \citealt{vanEerten11}), and $n_0$=of 0.01, 0.1 and 1 
cm$^{-3}$ (e.g.\ \citealt{Berger14}). Note that, since \toa\ and 
$\theta_j$ scale with $E_{52}/n_0$, some of these combinations are 
degenerate. 

\begin{figure}
\begin{center}
\psfig{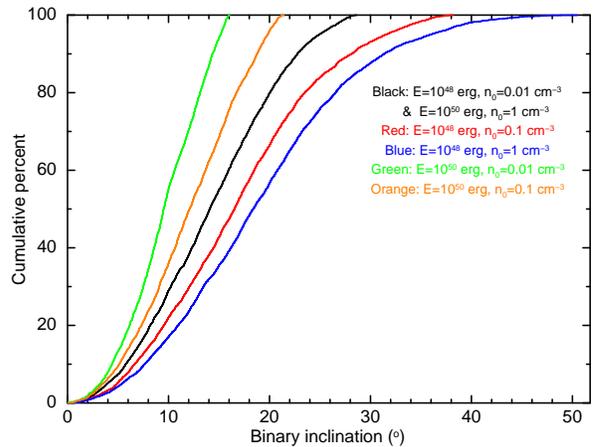}
\end{center}
\caption{The cumulative distribution of binary inclinations for the mergers which were detected
by our simulations, when the jet opening angle was 10\deg. The range of angles which result
in detectable GRBs depends strongly on the jet energy and circumburst density.}
\label{fig:inclination}
\end{figure}

In Fig.~\ref{fig:offaxis} we show the number of simulations in which the 
GRB was detectable in observations of 50 s or 500 s, as a function of 
the time at which the observation began. The on-axis and off-axis GRBs 
are differentiated. Although there are many more off-axis bursts (98.5\%\ of those simulated),
they are less detectable, since, by the time the 
emission becomes visible to the observer, it has faded significantly -- 
Fig.~\ref{fig:inclination} shows the cumulative distribution of binary 
inclinations for the GRBs detected and shows that this still corresponds 
to a fairly restrictive range of viewing angles. Fig.~\ref{fig:offaxis} 
shows that the detectability of the off-axis objects depends strongly on 
the jet energy and the circumburst density; in reality some (not 
currently constrained) distribution of these parameters is expected, 
which means that the `true' shape of Fig.~\ref{fig:offaxis} for the 
short GRB sample is some form of weighted average of the panels shown.

This variation in detectability makes it difficult to define the optimal 
time to observe in X-rays in order to detect off-axis emission. For 
on-axis GRBs, it is clear that the earlier the observations, the more 
GRBs are detected. For off-axis objects, observing too early 
prevents detection (the afterglow emission is still beamed away from the 
observer), as does observing too late (the emission is too faint), but 
the parameters of the window in which the afterglow can be detected vary 
according to the burst properties. Fig.~\ref{fig:offaxis} shows that by \til36 hours 
after the trigger, the number of off-axis bursts detectable exceeds the 
number of on-axis bursts for all $E_{52}=0.0001$ simulations. For if 
$E_{52}=0.01$ then on-axis bursts dominate until at least 72 hours post 
merger. A sensible \swift\ response to a GW trigger may therefore be to 
begin observations as soon as possible (in case the merger was on-axis), 
but after \til36 hours, if the afterglow has not been detected, the 
strategy should instead turn to looking for off-axis afterglows which 
are starting to emerge. In this case, rather than continuing the XRT 
observing program (which will be moving to lower-probability portions of 
the GW error region) it would be preferable to re-observe the 
high-probability fields already studied. This has an additional science 
bonus, in that if the GRB afterglow is detected in these second 
observations, but was not present in the first, it confirms that the 
merger was off-axis, and allows us to place constraints on $i-\theta_j$. 
One could envisage a third set of observations beginning at 72 hours after the
trigger (i.e.\ where the off-axis start to dominate over the on-axis GRBs if $E_{52}=0.01$).
However at this time, the number of off-axis afterglows detectable if $E_{52}=0.0001$ is still
much higher than if $E_{52}=0.01$, therefore unless there are compelling reasons to believe
that the true sGRB population is dominated by the higher-energy objects,
it is better to continue the observations than the return the the previously observed-fields.

We also see from Fig.~\ref{fig:offaxis} that the choice of exposure time
is much more important for off-axis GRBs than for on-axis GRBs, with the number of
off-axis GRBs detected a factor of 3--5 higher for 500 s exposures than for the 50 s exposures:
for on-axis GRBs the difference is at most a factor of 1.5. Therefore, if this 
`double-headed' approach were employed it would make sense to use shorter exposures in the
first 36 hours, while searching for on-axis objects, and then switch to longer exposures
when the search focus shifts to off-axis GRBs.

While the jet opening angle of 10\deg\ used above may be reasonably 
representative of typical jets \citep[e.g.][]{Duffell15}, lower limits 
as large as 25\deg\ have been deduced for sGRBs \citep{Grupe06,Zhang15}. We 
repeated our simulations for jet opening angles of 20\deg\ and 30\deg, 
extending the runs to 28 days after the merger, to account for the later 
rise time of the off-axis jets. Fig.~\ref{fig:offaxis_wide} shows the 
results for the values of $E_{52}/n_0$ which maximise the number of 
off-axis bursts detectable. Increasing the jet angle had two effects: 
some off-axis GRBs became on-axis GRBs, and the rise time of the 
off-axis GRBs moved to later times. The latter also means that the 
off-axis afterglows are fainter. These factors conspire to make the off-axis afterglow
less significant as a proportion of afterglows detectable with XRT.

All of the considerations discussed in this section 
show that chasing off-axis afterglows with a narrow-field
instrument such as \swift-XRT is extremely challenging. The number of such afterglows
that can be detected, and the time at which they are detectable, are both highly variable,
depending on the details of the jet and the medium in which burst occurs.

\section{Discussion and conclusions}
\label{sec:disc}

We have performed simulations of \swift-XRT follow-up of GW triggers 
from the ALIGO/AVirgo observatories, according to their expected 
performance in the 2015 and 2016 science runs. In each case, we assumed 
that ALIGO/AVirgo detected a NS-NS merger accompanied by an 
sGRB with an X-ray afterglow similar to those observed by \swift-XRT to 
date. We have found that XRT can identify the GRB in up to \til30\%\ of cases,
depending on the observational strategy employed. We considered this strategy in terms
of two questions: which portions of the GW error region to observe,
and how best to observe them.

For the former question, it is clear that in 2015 preferentially 
observing galaxies in the GW error region significantly increases the 
fraction of GRB afterglows (arising from NS-NS mergers) that we can identify with XRT. Beyond 2015
(or when considering NS-BH mergers), as the 
sensitivity of the GW network increases, galaxy catalogues that extend
to greater distances (200-300 Mpc) while remaining highly complete will be necessary.
One possibility is the 2MPZ catalogue \citep{Bilicki14}: its advantages are
large sky coverage and uniformity, its disadvantages
are use of photometric redshifts and the current lack of detailed
completeness analysis.

If \emph{Fermi-GBM} also triggers on the event detected by the GW 
facilities, we can convolve the GBM and GW error regions, which reduces 
the amount of sky we need to cover, typically by a factor of \til2.

Concerning the optimal strategy, our simulations show that the fraction of GW-detected sGRBs we can detect
with \swift-XRT does not strongly depend on the amount of time for which we observe
each field in our follow-up. Not surprisingly, the more fields are observed, the more
GRBs are identified, but this is a sub-linear relationship: doubling the number of fields
less than doubles the number of identified counterparts.

As we pointed out in Section~\ref{sec:simulate}, because we used the \swift\ X-ray afterglows
as templates in our simulations, our results are biased towards the brighter
sGRBs. Further, we have argued (Section~\ref{sec:howbright}) that sGRBs like those seen by \swift\
occur only extremely rarely within the distance that ALIGO/AVirgo will probe in the next few years,
suggesting that if a NS-NS merger is detected by these facilities, the accompanying sGRB will
either be (much) less luminous than those we have simulated, or will have its jet oriented away from us.
Indeed, if sGRBs follow a luminosity function such as that of \cite{Wanderman15}, a nearby GRB
will likely be low luminosity compared to the ones \swift\ has detected to date.
These considerations suggest that longer XRT exposures are preferable, to increase our
sensitivity to these fainter signals: in Section~\ref{sec:offaxis} we showed that the number
of off-axis  objects detectable increases by a factor of 3--5 when 500 s exposures are used, compared
to 50 s exposures. However, fainter afterglows will not be as easy to identify from among the uncatalogued 
unrelated sources also detected in longer exposures. In this case, to identify the GRB
requires a second observation, to search for evidence of fading. This observation will have
to be performed for every uncatalogued source detected, since we cannot know \emph{a priori} which (if any)
is the GRB. As the lower panel of Fig.~\ref{fig:nseren} shows, for the longer initial observations
(500 s per field), this may be many hundreds of sources. \cite{Evans15} showed that typically
an exposure of 1 ks would be enough to identify whether a source detected in these initial observations
had faded, with 3-$\sigma$ confidence, so a simplistic calculation\footnote{Neglecting, for example,
the possibility of observing multiple sources in a single field of view.} based on Fig.~\ref{fig:nseren}
suggests that the time needed to perform these follow-up observations 
is similar to the time taken to perform the initial observation of the GW error region. 

Even if we only perform short exposures with \swift, i.e.\ focus purely on the
bright but rare events which are detected above the RASS limit, some false positives
will still be detected simply because of Poisson effects. Therefore, it will be necessary
to re-observe even the bright, uncatalogued sources; however there are many fewer of these than
of the faint uncatalogued object just discussed so the required observing time is correspondingly lower.

If we do perform follow-up observations of any uncatalogued sources, and 
identify fading behaviour, we cannot rule out that the object is an 
unrelated variable object. As we noted in Section~\ref{sec:falsePos}, 
roughly \til1/72 sources unrelated to the GW event are expected to show 
strong signs of fading, therefore if a fading source is found, continued 
observations will be necessary to definitively identify it as the GRB 
afterglow.

The discussion so far has focussed on the attempt to identify the 
counterpart to the GW trigger based on \swift\ data alone, whereas a 
large number of facilities spanning the EM spectrum have committed to 
the follow up of GW triggers. While X-rays may represent the best 
wavelength at which to rapidly detect an on-axis sGRB afterglow, sGRB 
emission is panchromatic, and for off-axis events the optical and 
infrared emission from a kilonova represents arguably the best hopes for 
a detection (see \citealt{Metzger12}). Kilonova emission is powered by 
the radioactive decay of r-process elements created during the merger 
\citep{Rosswog99,Freiburghaus99}, and detection of such emission has 
been claimed for GRB 130603B \citep{Tanvir13a,Berger13}, as well as 
possibly GRB 080503 \citep{Perley09,Gao15} and GRB 060614 
\citep{Yang15}. The knowledge of what other observatories have detected 
-- or have not detected -- will therefore be of significant value in 
trying to establish which, if any, of the X-ray sources detected is the 
GW counterpart. However, signals such as the kilonova may not rise 
until some time after the trigger in which case these data would not be 
available to us at the time of the \swift\ observations, so the ability 
to identify the counterpart from \swift\ data alone is desirable.

Due simply to geometric effects, off-axis mergers are much more likely than on-axis ones: for a 10\deg\ jet,
only 1.5\%\ of mergers are on-axis. However, a much lower fraction of off-axis objects are detectable
by XRT, and only within a specific window of time after the merger event. The timing 
of this window depends strongly the jet opening angle, the energy in the jet, and the
density of the circumburst medium. The distributions of these parameters for sGRBs
are not well known, therefore it is difficult to identify the `best' time to observe
when searching for an off-axis afterglow. This in particular highlights the fact that the ideal
facility would be a wide-field (preferably all-sky) X-ray imager with good position accuracy 
\citep[e.g.][]{Osborne13}.

As \swift-XRT has a narrow field of view, we require many pointings to observe the GW error region,
and new flight software to support the upload of hundreds of `Automatic Targets' is being developed to support this.
We have simulated some fairly extreme observing programmes: for example, having \swift\ 
observe 1,800 fields for just 10 s each; whether this is practical is currently unclear and,
at the time of writing, the maximum density of observations with \swift\ is still under study, 
However, we do know that 
the most slews \swift\ has performed in a single day is 167. This corresponds to just over 500 s per 
observation, if we ignore the time spent slewing or passing through the South Atlantic Anomaly (when XRT cannot
collect data). Therefore, our 500-s per observation simulations give us a reasonable idea
of what we can expect in reality. In this regime, if we observed for 1 day (i.e.\ \til 150 fields)
Fig.~\ref{fig:nObs_effect} shows us that, if the NS-NS merger were on-axis, we would identify 
the sGRB in up to \til18\%\ of cases in 2015, and up to \til25\%\ in 2016.
Observing for 4 days (600 fields) this becomes 25\%\ in 2015 and \til 32\%\ in 2016. These numbers assume
that nearby sGRBs will be comparible in luminosity to those detected to date, which is unlikely
therefore they represent upper limits on the fraction we can identify.

In conclusion: the X-ray wavelength offers a promising means to identify 
the EM counterpart to a GW trigger, due to the low rate of bright 
transients compared to other wavelengths, and the fact that X-ray 
facilities are  in orbit, and thus not subject to 
problems of daytime. \swift, with its flexible planning and rapid 
slewing abilities, is able to cover significant portions of the sky, and 
by conducting a series of short (50--500 s) exposures, has a reasonable chance of 
identifying any X-ray counterpart to the GW trigger that may exist.

\appendix
\section{The XRT Simulations}
\label{app:sims}

To simulate the actual follow-up observations, we used the following
approach (when we describe a field as `observable' below, we mean
that it is observable by \swift-XRT for at least $e$ s):

\begin{enumerate}

\item{Set a counter $I=1$. This keeps track of how many iterations over
all $N$ fields have been attempted. Set the time to $\tau_d$ (Section~\ref{sec:simulate}).}

\item{Identify the most probable field observable. If no field is
observable, identify which one will become so soonest, and wait until
then.}

\item{\label{step:observe}Increase the exposure of the current field by
$e$ s.}

\item{For each unrelated source in the field which has not yet been
detected, determine the probability that the source is now detected,
based on the its count-rate, the exposure in this field so far, and the
completeness of the detection system (\citealt{Evans14}, section~7 and fig.~14). Then
generate a random number between 0 and 1 to determine whether the source
is detected.}

\item{If the GRB lies in the field, determine how many counts are
detected from it in this exposure, from the redshift-corrected afterglow
template defined in Section~\ref{sec:simulate}. Add this to any counts detected
in previous (or overlapping) observations of the GRB. Then determine, whether or not it is detected,
as for the unrelated sources. If it is detected, determine the mean
count-rate of the detection, and compare this to the RASS upper limit
for this field. Also, construct a light curve of the GRB, and determine
whether fading can be determined with 3-$\sigma$ confidence. If the GRB
is above the RASS limit, or fading, mark it as `identified' and record
the time.}

\item{If all fields that are not too close to the Sun, Moon or orbital
pole to observe, have been observed $I$ times, increase $I$.}

\item{If there are any fields which overlap the current field, are
observable, and have been observed $<I$ times, identify the one with the
greatest probability, and select it.\label{step:nextfield}}

\item{If there are no overlapping fields matching the above criteria,
identify all currently observable fields that have been observed fewer
than $I$ times, and select the one with the highest probability. If any
field which has been observed fewer than $I-1$ is observable (e.g.\ it was Sun constrained during the earlier rounds of
observations, but is now observable), select this field in preference to
the other fields, regardless of probability.}

\item{If no fields are currently observable, determine when the next
field that has been observed fewer than $I$ times will become visible,
and wait until this point, then select that field.}

\item{If the time is less than 15 days since the trigger, go to
step~\ref{step:observe}, otherwise, end the simulation.}

\end{enumerate}

\noindent Step~\ref{step:nextfield} exists for efficiency reasons: by observing
overlapping fields consecutively by preference, the ratio of time spent
observing to time spent slewing is maximised. The GW probability maps
have smooth contours, therefore the only time this step results in
selecting a field with lower probability than a non-overlapping field is
when the map bifurcates, in which case not including
step~\ref{step:nextfield} would tend to result in \swift\ constantly
slewing between the two disjointed regions, wasting much valuable time.

The simulations assume that to move from a field to an overlapping 
one requires 25 s of `dead time' between the end of one observation and 
the beginning of the next. This value was determined by examining data 
from the \swift\ tiled observations used to respond to neutrino
triggers \citep{Evans15}. The situation when performing a slew between
disjointed fields is more complex, as it depends on various details of
the slew, which is beyond the scope of this work to simulate. Based
on an analysis of historical \swift\ observations, we assumed a slew rate of 
0.77\deg\ s$^{-1}$, plus 30 s of overhead.

After each observation, the simulation records the time since the
trigger, how many unrelated sources have been detected so far, and
how many of those are not in the RASS, and whether or not the GRB has
been detected, and identified.

\section*{Acknowledgements}

This work made use of data supplied by the UK Swift Science Data Centre
at the University of Leicester, and used the ALICE High Performance
Computing Facility at the University of Leicester. PAE and JPO acknowledge
UK Space Agency support. SC is supported by ASI-INAF Contract I/004/11/1

\bibliographystyle{mn2e} \bibliography{phil}

\label{lastpage} \end{document}